\documentclass[baaa]{baaa}

\usepackage[pdftex]{hyperref}
\usepackage{subfigure}
\usepackage{natbib}
\usepackage{helvet,soul}
\usepackage[font=small]{caption}

\contriblanguage{1}

\contribtype{2}

\thematicarea{10}

\title{
Archaeoastronomical study of Christian churches \\ in Fuerteventura
}

\titlerunning{Archeoastronomy in Fuerteventura}

\author{
M.F. Muratore\inst{1,2}
\&
A. Gangui\inst{1,3}
}

\authorrunning{Muratore \& Gangui}

\contact{flormuratore@gmail.com}

\institute{
Instituto de Astronomía y Física del Espacio, CONICET--UBA, Argentina
\and
Departamento de Ciencias Básicas, UNLu, Argentina
\and
Facultad de Ciencias Exactas y Naturales, UBA, Argentina
}

\resumen{
Presentamos un estudio arqueoastronómico de las orientaciones de las iglesias cristianas coloniales de la isla de Fuerteventura, en Canarias, España, construidas en su mayoría desde el período de la conquista normanda en el siglo XV hasta el siglo XIX. Nuestro objetivo es analizar la posible influencia astronómica en la orientación de estas iglesias. Los resultados preliminares sugieren que la gran mayoría de las construcciones religiosas de la isla posee sus ejes orientados dentro del rango solar, entre los acimuts extremos del movimiento anual del Sol al cruzar el horizonte local. Esto difiere de lo hallado en las islas de Lanzarote y La Gomera (también en Canarias) previamente estudiadas.
}

\abstract{
We present an archaeoastronomical study of the orientations of the colonial Christian churches on the island of Fuerteventura, in the Canary Islands, Spain, mostly built from the period of the Norman conquest in the 15th century to the 19th century. Our goal is to analyze the possible astronomical influence on the orientation of these churches. Preliminary results suggest that the vast majority of the island's religious constructions have their axes oriented within the solar range, between the extreme azimuths of the annual movement of the Sun as it crosses the local horizon. This differs from what was found on the islands of Lanzarote and La Gomera (also in the Canaries) previously studied.
}


\begin{document}

\maketitle

\section{Introduction}\label{S_intro}

From the very beginning, Archaeoastronomy has mainly focused on the study of historical constructions, such as temples, megalithic monuments and structures built by different cultures around the world, and on the analysis of the possible influence of celestial objects and phenomena on their design \citep{ruggles_2015}.
In this context, one of the most relevant aspects to approach in this field is the study of the orientation of the constructions, which might provide valuable information regarding the intentionality of the people who built them. 
In particular, we are interested in studying the orientation of Christian churches in Fuerteventura, one of the Canary Islands in Spain, from the time of the Norman conquest in the fifteenth century to the nineteenth century.
Ancient texts state that Christian churches should be oriented with their apse towards the east. In other words, their main axis, that is, the line from the entrance to the altar, should point in the direction from where the Sun rises in a particular day of the year, in this case, the equinox \citep{mccluskey_1998}. 
In this work we present a brief summary of our analysis and very preliminary results obtained from data previously collected by our research group to test if this is the general case in Fuerteventura \citep{muratore_gangui_2020}.
Our goal is to investigate if the data supports a possible relationship between the churches orientations and astronomical phenomena. 
As a future project, we also aim to investigate whether the pre-Hispanic population already existing on the island at the time of the conquest might have had influence on the construction 
of the first Christian churches. This research is part of a larger project that includes previous work on other islands of the same region \citep{gangui_lanzarote_2016,di_paolo_la_gomera_2020}.

\section{Data and methodology}

In order to analyze the orientation of the churches, on-site measurements taken in a previous field work were used.
These consist of the azimuth and the altitude of the horizon corresponding to the direction of the main axis of each church.
In those cases where the elevation was difficult to determine, for instance due to the presence of modern buildings, the on-line tools available at heywhatsthat.com allowed us to reconstruct the horizon using a digital model of each site. 

The values of the azimuths, measured using a high precision compass, were then corrected by magnetic declination, allowing us to achieve a precision of $0.5^\circ$ in the astronomical azimuths. These directions are 
in general indicated in circular azimuth diagrams as the one shown in Fig. 5 of \cite{muratore_gangui_2020}, in order to help visualize the orientations obtained.

Given that there are 48 churches distributed in approximately $1700~\mathrm{km}^{2}$, we consider the sample to be representative and significant for this type of study.

\begin{figure}[!t]
\centering
\includegraphics[width=\columnwidth]{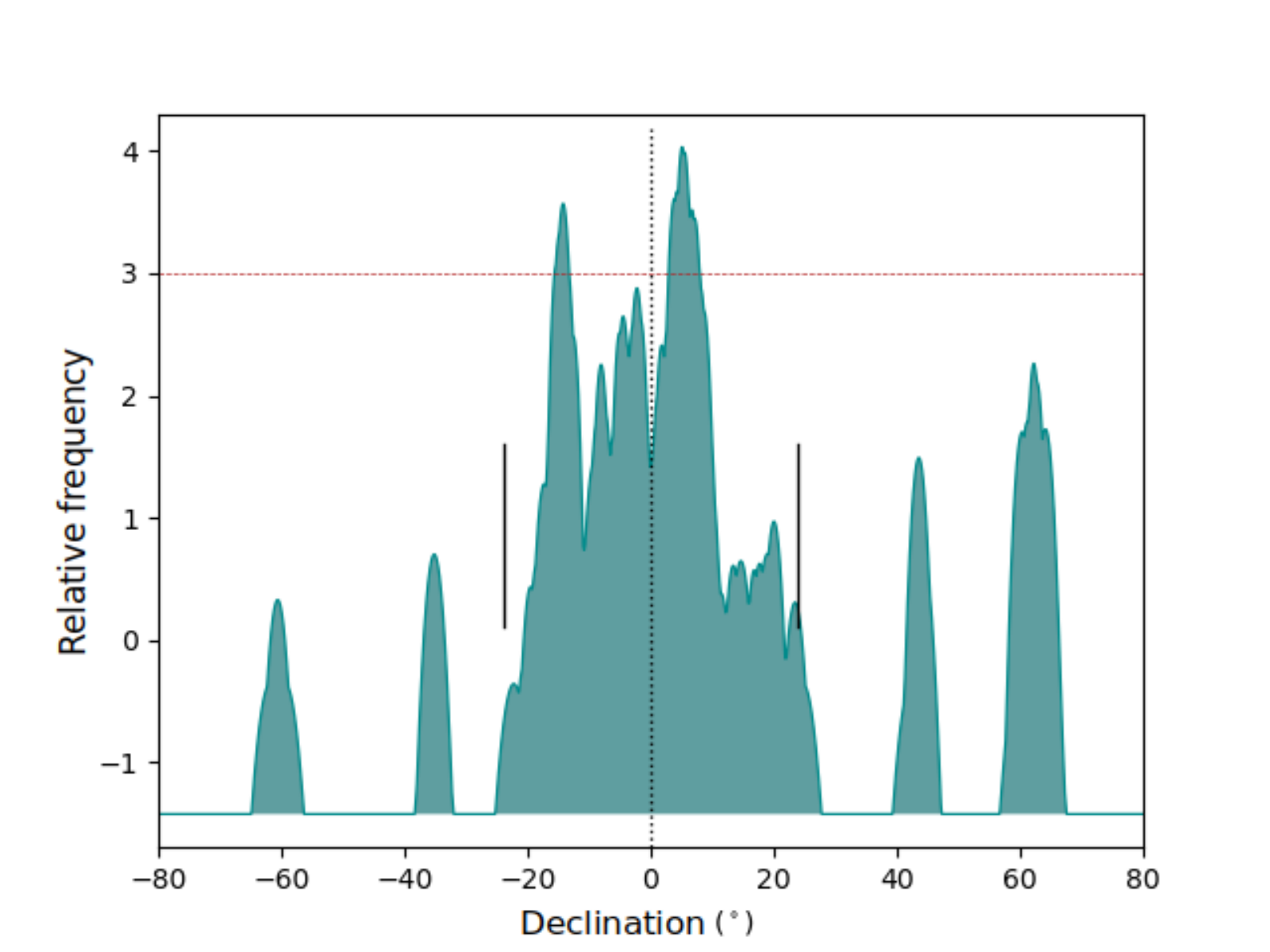}
\caption{Declination histogram for the colonial Christian churches in Fuerteventura. The plotted data was estimated from the values of the azimuths and heights of the horizon at each site obtained from the on-site measurements \citep{muratore_gangui_2020}, and corrected by the effect of local magnetic declination and atmospheric refraction. A couple of statistically significant peaks are found above the 3$\sigma$ level (horizontal dotted line). 
}
\label{Figura}
\end{figure}

In order to have a measure of the orientations which is independent of the geographical location and topography across the island, the azimuth and the altitude, together with the latitude of each site under study, are used to determine the astronomical declination corresponding to the direction of the main axis of the church. 
The declination histogram for the whole group of constructions is given in Fig.~\ref{Figura}.

The quantities thus obtained may then be compared with the declination of the Sun for different days around the date of construction of each particular temple. This will allow us to investigate if there is a possible connection between the Sun and the architecture of the structure, which in turn might help us to find a possible explanation for the orientation of the church. A thorough statistical analysis of the declination distribution that we found is currently in progress, as it is necessary to confirm the general pattern that was observed in the azimuth diagram.

\section{Results}

Fig.~\ref{Figura} shows a clear accumulation of declination values within the solar range (delimited by the short vertical bars in the picture), supporting the trend already found in the azimuthal orientation of the churches in Fuerteventura \citep{muratore_gangui_2020}. While some of the constructions do not follow this trend, 35 out of the 48 churches do show orientations that fall within the solar range. Moreover, most of them point towards directions not far from east, corresponding to the rise of the Sun near the equinoxes. This is now confirmed by our declination histogram, where we see that the frequency of orientations increases remarkably between the extreme positions of the Sun at the solstices. 
This result is consistent with the pattern of orientations found in historical churches built before the conquest in the regions where the colonizers were originally from.

On the other hand, some of the temples in our sample present orientations outside the solar range. This may be due to different reasons. For example, some historical churches were built as private temples by landowners, which means that their position and orientation  were subject to the distribution of the existing buildings. In a few cases the construction of the structures was limited by, or planned according to, the topography of the island, including coastlines, gorges and mountains. In the case of some modern churches built within big cities, there was probably no possibility or intention to choose their orientation according  to the early Christian texts. \\

\section{Discussion}

As mentioned in the Introduction, according to early prescriptions the apse of Christian churches should face east, that is, point towards the rising Sun during equinoxes. Since the definition and determination of the date corresponding to the equinox may be affected by different factors \citep{ruggles_1997}, this kind of study must include not only the reconstruction of the sky in the epoch when the construction was planned, but also the possible influence of the equinox used as reference.

On the other hand, one of the main difficulties regarding the knowledge about the churches is to find reliable information about them, and sometimes, no information can be found at all. For example, to delimit the epoch when they  were planned or when their construction began the information is not always easy to find, and in some cases, different sources make reference to different years or even different centuries. A detailed analysis of historical documents from different periods since the colonization, such as letters, official and informal reports, and records kept by the people in charge of the churches themselves will be extremely important. The search for other sources of information will also be necessary. 
 
The uncertainty in the date of construction of the churches actually leads to a range  of possible declinations for the Sun, making it difficult to establish if there is a correlation, for example, between the orientations and the dates of the equinoxes and solstices.

Once the dates are determined, the change from the Julian to the Gregorian Calendar introduced in 1582 has to be taken into account, not only for establishing the epoch when the construction was planned, but also for determining the dates, for example, of the equinox corresponding to that particular epoch. 

The influence of the original island inhabitants on the orientation of the first temples might also be expected, since this is the case in other parts of the Canarian Archipelago \citep{belmonte_2015}. 
However, pre-Hispanic orientations would be mainly solstitial,  while our first results seem to indicate  that most of the temples face towards directions relatively close to the equinox.
During the first years after the conquest of Fuerteventura, the original inhabitants and the colonists  gave rise to a new society, where both  coexisted. 
This might have resulted in a majority of temples oriented within the solar range (although not always strictly towards the equinox), with some of the churches orientations corresponding to important dates for the community, which includes both groups, each one with its own religious customs.

On the other hand, a thorough orographic study of the island has to be carried out. A complete analysis must include detailed topographic maps, which will allow us to study  the geographical profile and the altitude of the horizon around the temples. The direction of strong winds also has to be taken into account, since it could influence the orientation of the churches in some parts of the island.

Let us note that in La Gomera and in Lanzarote, two of the islands of the Canarian Archipelago where the orientations of the churches and their possible origin were thoroughly studied, we find a completely different scenario than the one suggested by the preliminary results obtained in Fuerteventura.

Even though further studies might reveal new information that leads to alternative interpretations, the results obtained so far for the orientations of the historical Christian churches of 
Fuerteventura strongly suggest that in this island, the prescriptions for canonical orientations were followed.

Future work will include the study and comparison of these results with those obtained for other islands, as well as with the orientations of historical Christian churches built in the European continent during the same period as the ones built in the Archipelago.

\begin{acknowledgement}


The authors wish to thank their collaborators Maitane Urrutia, Juan Belmonte, Carmelo Cabrera and César González-García for many useful discussions during the data analysis and for their support during the field work. This work has been partially financed by CONICET, Universidad de Buenos Aires, Universidad Nacional de Luján and by the projects P/309307 Arqueoastronomía, from Instituto de Astrofísica de Canarias, and Orientatio ad sidera IV - AYA2011-66787-P supported by Spain's MINECO. M.F.M. has a CONICET doctoral fellowship.

\end{acknowledgement}


\bibliographystyle{baaa}
\small
\bibliography{ID_654}

\begin{thebibliography}{}

\bibitem[\protect\citeauthoryear{{Belmonte}}{{Belmonte}}{2015}]{belmonte_2015}{Belmonte} J.A., 2015, {Handbook of Archaeoastronomy and Ethnoastronomy}, 1115-1124, Springer-Verlag New York, New York
\bibitem[\protect\citeauthoryear{{Di Paolo}}{{Di Paolo} et~al.}{2020}]{di_paolo_la_gomera_2020}{Di Paolo} A., et~al., 2020, Cosmovisiones/Cosmovis\~oes, 1, 73
\bibitem[\protect\citeauthoryear{{Gangui}, {González García}, {Perera Betancort} \& {Belmonte}}{{Gangui} et~al.}{2016}]{gangui_lanzarote_2016}{Gangui} A., et~al., 2016, Tabona, 20, 105
\bibitem[\protect\citeauthoryear{{McCluskey}}{{McCluskey}}{1998}]{mccluskey_1998}{McCluskey} S.C., 1998, {Astronomies and cultures in early Medieval Europe}, Cambridge University Press, Cambridge
\bibitem[\protect\citeauthoryear{{Muratore} \& {Gangui}}{{Muratore} \& {Gangui}}{2020}]{muratore_gangui_2020}{Muratore} M.F., {Gangui} A., 2020, Anales AFA, 31, 93
\bibitem[\protect\citeauthoryear{{Ruggles}}{{Ruggles}}{1997}]{ruggles_1997}{Ruggles} C.L.N., 1997, Journal for the History of Astronomy Supplement, 28, S45
\bibitem[\protect\citeauthoryear{{Ruggles}}{{Ruggles}}{2015}]{ruggles_2015}{Ruggles} C.L.N., 2015, {Handbook of Archaeoastronomy and Ethnoastronomy}, Springer-Verlag New York, New York

\end{thebibliography}
 
\end{document}